\documentclass[conference,a4paper]{IEEEtran}
\usepackage{amsmath}
\usepackage{enumerate}
\usepackage[dvips]{graphicx} % for figures
\usepackage{amsfonts}
\usepackage{amssymb}
\usepackage{amscd}
\usepackage{subfigure}
\usepackage{xcolor}

\begin{document}

\sloppy

%% Paper Title
\title{Easily Implemented Rate Compatible Reconciliation \\Protocol for Quantum Key Distribution}

\author{
  \IEEEauthorblockN{Zhengchao Wei and Zhi Ma}
  \IEEEauthorblockA{State key Laboratory of Mathematical Engineering and Advanced Computing, Zhengzhou\\
Email:yhweizhengchao@gmail.com}}

%% To balance the two columns, you should reduce the text-height of
%% the last page using the following command:
%%%%%%%%%%%%%%%%%%%%%%%%%%%%%%%%%%%%%%%%%%%%%%%%%%%%%%%%%%%%%%%%%%%%%
%\addtolength{\textheight}{-9.35cm}
%%%%%%%%%%%%%%%%%%%%%%%%%%%%%%%%%%%%%%%%%%%%%%%%%%%%%%%%%%%%%%%%%%%%%
%% with an appropriate value. This command must be place on the second
%% last page, i.e., for a one-page abstract here, for a two-page
%% abstract right after the \maketitle command.

%% Create the title:
\maketitle

\begin{abstract}
  Reconciliation is an important step to correct errors in Quantum Key Distribution (QKD). In QKD, after comparing basis, two legitimate parties possess two correlative keys which have some differences and they could obtain identical keys through reconciliation.

  In this paper, we present a new rate compatible reconciliation scheme based on Row Combining with Edge Variation (RCEV) Low Density Parity Check (LDPC) codes which could change code rate adaptively in noisy channel where error rate may change with time. Our scheme is easy to implement and could get good efficiency compared to existing schemes. Meanwhile, due to the inherent structure we use, the new scheme not only saves memory space remarkably but also simplifies the decoder architecture and accelerates the decoding.

\end{abstract}

\section{Introduction}

QKD is an important application of quantum information science, and it is unconditionally secure in theory even if there is an eavesdropper who has infinite computing ability. QKD protocol aims to make two legitimate parties share a identical key string. The best known protocol of QKD is BB84 which was first proposed in 1984~\cite{bennett1984quantum}. Reconciliation is one of the steps in post-processing, the other one is called privacy amplification which is used to reduce the information obtained by eavesdropper. Reconciliation is carried out after two participators who we call Alice and Bob for convenience comparing their chosen basis through an authenticated classical channel and possessing two correlated strings with some differences. Then they use reconciliation to eliminate the differences in their strings.

The efficiency of reconciliation $f$ can be defined as:
\begin{equation} \label{introduction:f1}
\begin{aligned}
f = \frac{{I\left( e \right)}}{{H\left( e \right)}}
\end{aligned}
\end{equation}
where $I\left( e \right)$ represents the information consumed to reconcile in practice, and $H\left( e \right)$ is the consumed information in ideal condition. $f$ is a function of error rate which is larger than $1$ and the efficiency of scheme is better if it is closer to $1$. In binary symmetric channel with reconciliation protocols based on LDPC codes which we concern in this paper, $f$ can be written as:
\begin{equation} \label{introduction:f2}
\begin{aligned}
f = \frac{{1 - R}}{{H(e)}}
\end{aligned}
\end{equation}
where $R$ is the code rate of LDPC codes,  $e$ represents channel error rate, and
$H\left( e \right) =  - e{\log _2}e - \left( {1 - e} \right){\log _2}\left( {1 - e} \right)$ is the binary entropy function.

The authors of BB84 proposed reconciliation scheme BBBSS~\cite{bennett1992experimental} to correct errors. BBBSS is based on binary search to find errors. In BBBSS, Alice and Bob need to implement protocol several rounds to make the errors in key strings few enough. In each round, Alice and Bob permute their strings according to a common permutation. Then they divide their strings into several blocks and test whether the parities of corresponding blocks are equal. If the parities do not agree, there must be an odd number of errors in corresponding blocks, then Alice and Bob use binary search to find one error. Executing sufficient rounds can guarantee the errors in strings few enough. Because BBBSS can not correct errors efficiently when there are an even number of errors, Brassard and Salvail presented Cascade protocol \cite{brassard1994secret}. Cascade is also based on binary search to correct errors. However, in contrast to BBBSS, in Cascade, when Alice and Bob find an error in a new round, they know there must be an odd number of errors in the blocks which containin this error bit in earlier rounds, and then they correct errors in previous rounds by binary search. Alice and Bob repeat this process until there is no new block with an odd number of errors.

Although Cascade can get good efficiency, it should be noted that both protocols need a lot of communication, which will put extra pressure on QKD. Pearson proposed a scheme~\cite{pearson2004high} based on LDPC codes which was first introduced by Gallager in 1962~\cite{gallager1962low}. In this scheme, Alice just sends the syndrome of her keys which is the compressed information of her keys to Bob and Bob could decode his keys with the syndrome obtaining from Alice and tell Alice whether the protocol succeeds or not. Therefore the interactivity is low in this protocol.
In real life, the error rate of channel may change with time, so an LDPC code is practical if it could change its code rate adaptively to fit the error rate. Two commonly used methods to achieve this are puncturing and shortening~\cite{tian2005construction}. To improve the adaptability, some schemes based on the puncturing and shortening method were proposed by Elkouss and Martinez et al~\cite{elkouss2010rate,martinez2010interactive,martinez2012blind}. In these protocols, puncturing improves the code rate by sender Alice setting some positions to be random numbers which are unknown to receiver Bob and shortening decreases the code rate by Alice revealing some positions to Bob. In \cite{elkouss2010rate}, Alice and Bob use a part of their keys to estimate the error rate of channel, then they could choose a suitable LDPC code to realize reconciliation. In \cite{martinez2010interactive,martinez2012blind}, the estimation process was removed with the price of increasing interactivity. We should note that, the positions selected to be punctured and shortened should be chosen carefully which may slow down the reconciliation. Besides, puncturing will consume the random keys possessed by Alice and Bob by setting some positions of codeword to be random numbers.

In this paper, we propose a new rate compatible reconciliation scheme based on RCEV LDPC codes. The rest of the paper is organized as follow: in part \textbf{II}, we introduce the main idea of RCEV and the new scheme is presented. In part \textbf{III}, the simulation results are given.

\section{New Information Reconciliation Protocol}
\subsection{RCEV LDPC Codes}
An LDPC code could be represented by a Tanner graph. There are two sets in a Tanner graph. One is the set of variable nodes which correspond to the columns of check matrix, and the other is the set of check nodes corresponding to the rows of check matrix. A variable node is degree $i$ if it is connected to $i$ check nodes, and vice versa. The degree distribution of variable nodes and check nodes can be defined as:
\begin{equation} \label{degree distribution}
\begin{aligned}
\lambda \left( x \right) = \sum\limits_{i = 2}^{{d_v}} {{\lambda _i}{x^{i - 1}}} \\
\rho \left( x \right) = \sum\limits_{i = 2}^{{d_c}} {{\rho _i}{x^{i - 1}}}
\end{aligned}
\end{equation}
where $d_v$ is the maximum degree of variable nodes, and $d_c$ is the maximum degree of check nodes; $\lambda_i$($\rho_i$) is the proportion of edges which are connected to degree $i$ variable(check) nodes. Then the code rate can be written as:
\begin{equation} \label{keyrate degree distribution}
\begin{aligned}
R = 1 - \frac{{\sum\limits_{i = 2}^{{d_c}} {\frac{{{\rho _i}}}{i}} }}{{\sum\limits_{i = 2}^{{d_v}} {\frac{{{\lambda _i}}}{i}} }}
\end{aligned}
\end{equation}

\begin{figure}[hbt]
\centering
\includegraphics[width=8cm,height=6cm]{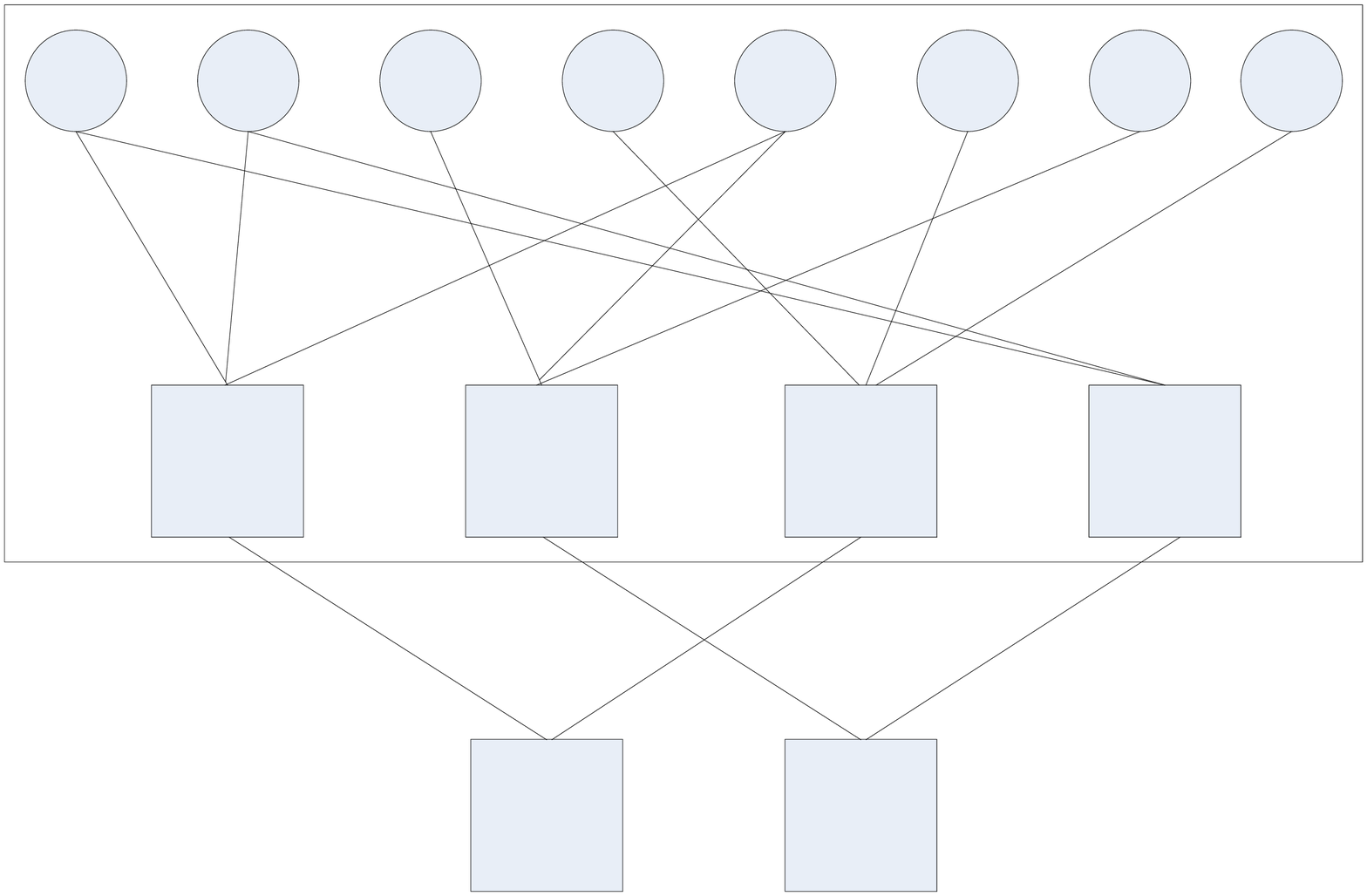}
\caption{The tanner graph of mother code and effective code. The circles represent the variable nodes and the square are check nodes. The code rate of above code (in box) is $1/2$ and the code rate of below code is $3/4$.}
\label{Fig:tanner}
\end{figure}

Here, we introduce the main idea of RCEV method. RCEV method derives from the row combining method which was first introduced by Casado et al. in \cite{casado2004multiple}. With row combining method, one could change a $[n,k]$ code to a $[n,k^{'}]$ code, $k^{'} > k$.

suppose there is a check matrix of LDPC code as:
\begin{equation}
\begin{aligned}
H=
\begin{pmatrix}
1&1&0&0&1&0&0&0\\
0&0&1&0&1&0&1&0\\
0&0&0&1&0&1&0&1\\
1&1&0&0&0&0&0&0
\end{pmatrix}
\end{aligned}
\end{equation}
We use this check matrix which is definitely not practical to illustrate the idea of row combining. To increase the code rate of this code, one could combine the rows of the check matrix. For example, to get a higher rate code, one could combine the first row with the third row, the second row with the fourth row, and the resulting check matrix is:
\begin{equation}
\begin{aligned}
H^{'}=
\begin{pmatrix}
1&1&0&1&1&1&0&1\\
1&1&1&0&1&0&1&0
\end{pmatrix}
\end{aligned}
\end{equation}
The code rate of $H$ is $1/2$ and the code rate of $H^{'}$ is $3/4$, so one can obtain a new code with higher code rate by row combining method. Row combining method could be represented by Tanner graph as showed in Fig.~\ref{Fig:tanner}. We call the code designed to generate higher code rate codes as mother code, and the generated codes as effective codes. We should note that the rows which have ``1'' in the same column in mother matrix should not be combined, i.e., the check nodes which share a common variable node in Tanner graph should not be combined. This is because if these rows were combined, the corresponding column is ``0'' in the new row and the information ``1'' provided in original rows is eliminated. To save the memory space and improve the efficiency of decoding, the mother matrix of our LDPC codes is designed to have inherent structure. In this structure, mother matrix is consist of a number of square sub-matrices with size $p$. Each sub-matrix is either a zero sub-matrix or a cyclically shifted identity sub-matrix. We use ``-1'' and ``i''to represent the zero-matrix and the sub-matrix produced by cyclically shifting $i$ places identity matrix. The matrix composed by these symbols is called base matrix. This structure could save memory space by only saving the base matrix and reduce the decoder complexity as showed in\cite{hocevar2004reduced,mansour2002low}.

A sub-matrix with size $5$ and a base matrix are given for example below:
\begin{equation}
\begin{aligned}
S_{3}=
\begin{pmatrix}
0&0&0&1&0\\
0&0&0&0&1\\
1&0&0&0&0\\
0&1&0&0&0\\
0&0&1&0&0
\end{pmatrix}
\end{aligned}
\end{equation}
$S_{3}$ is produced by cyclically shifting columns of $5\times5$ identity matrix $3$ places.
\begin{equation}
\begin{aligned}
H_b=
\begin{pmatrix}
{ - 1}&0&2&{ - 1}&{ - 1}\\
2&{ - 1}&0&{ - 1}&0\\
1&{ - 1}&{ - 1}&0&2\\
2&1&0&0&{ - 1}
\end{pmatrix}
\end{aligned}
\end{equation}
$H_{b}$ is a base matrix, and the size of extended matrix is $4p\times5p$, where $p$ is the size of sub-matrix. $H_{b}$ could be extended to $H^{''}$ with $3\times3$ sub-matrix:
\begin{equation}
\begin{aligned}
H^{''}=
\begin{pmatrix}
\begin{smallmatrix}
0&0&0&1&0&0&0&0&1&0&0&0&0&0&0\\
0&0&0&0&1&0&1&0&0&0&0&0&0&0&0\\
0&0&0&0&0&1&0&1&0&0&0&0&0&0&0\\
0&0&1&0&0&0&1&0&0&0&0&0&1&0&0\\
0&1&0&0&0&0&0&1&0&0&0&0&0&1&0\\
1&0&0&0&0&0&0&0&1&0&0&0&0&0&1\\
0&1&0&0&0&0&0&0&0&1&0&0&0&0&1\\
0&0&1&0&0&0&0&0&0&0&1&0&0&1&0\\
1&0&0&0&0&0&0&0&0&0&0&1&1&0&0\\
0&0&1&0&1&0&1&0&0&1&0&0&0&0&0\\
0&1&0&0&0&1&0&1&0&0&1&0&0&0&0\\
1&0&0&1&0&0&0&0&1&0&0&1&0&0&0
\end{smallmatrix}
\end{pmatrix}
\end{aligned}
\end{equation}

Row combining is equivalent to combining the check nodes which have no common variable nodes, and this results in the blocklength and the variable nodes degree distribution of mother code and effective codes maintain constant. Constant blocklength could simplify the decoder architecture. However, constant degree distribution may put a strict limit to the degree distribution and the performance of LDPC codes~\cite{casado2009multiple}. In fact, if the number of degree-two nodes exceeds the number of check nodes, there will be circles that consist of degree-two variable nodes and check nodes, which will decrease the performance of LDPC codes~\cite{tian2004selective}. These cycles become smaller as the number of degree-two nodes increase, which make the performance worse. Thus, with strict row combining method, the mother code is constructed with degree distribution which is optimal for the highest rate code to ensure the number of degree-two nodes will not exceed the number of check nodes in the highest rate code. This means that the number of degree-two nodes and the maximum degree of variable node are optimal for the highest rate code and they are generally smaller comparing to those in the degree distributions which are optimal for lower rate codes. Lower rate codes generally need more degree-two nodes to improve the performance, and larger maximum variable node degree results in a better code~\cite{richardson2001design}. As stated above, strict row combining method puts a limit on the performance of low rate codes.

To solve these problems, Casado et al. proposed Row Combining with Edge Variation (RCEV) codes in \cite{casado2009multiple}. The main idea of RCEV codes is to delete or increase edges when implementing row combining. With this method, mother code and effective codes could have different degree distribution. In RCEV method, the degree distribution of mother code is set to be optimal for lowest-rate code. To maintain the number of degree-two nodes less than the number of check nodes, some of degree-two nodes in lower rate code are translated to degree-three nodes in higher rate codes by translating a zero sub-matrix to a non-zero sub-matrix in check matrix, i.e., increasing an edge to some degree-two variable nodes. When two non-zero sub-matrices are combined, if the degree of combined variable nodes is not maximum variable node degree, the mother matrix should be regenerated, otherwise, we discard one of the combined sub-matrices. With this method, the inherent structure of mother matrix is kept. This structure makes RCEV codes can be decoded with Layered Belief Propagation (LBP) method, which is introduced in \cite{mansour2003high,hocevar2004reduced} and could accelerate the decoding by using nearly half iterations. The performance of mother code and effective codes is guaranteed by ACE detection \cite{tian2004selective} which improves the performance of short cycles.

\subsection{Protocol}
We propose a reconciliation scheme based on RCEV LDPC codes here. Initially, after sending and measuring pulses through quantum channel, Alice and Bob compare the basis used to encode and decode pulses through classic channel. Then they maintain the the keys extracted from the pulses where they use the same basis. Assume that Alice and Bob possess key strings $X$ and $Y$ respectively. There are a few of discrepancies between $X$ and $Y$, and they need to reconcile these two strings through classic channel. Alice and Bob estimate the region of the classic channel error rate by experience and choose a proper variable node degree distribution corresponding to the lowest error rate. Then they generate the mother code and effective codes by RCEV method. Alice and Bob store the base matrix of mother code and the strategies of RCEV in memory. With the base matrix and strategies, Alice and Bob could then generate codes with different code rates which could cover the region of the classic channel error rate. Once these preparation work is finished, Alice and Bob reconcile key strings as follows:
\begin{enumerate}[step 1:]
\item
Alice and Bob estimate the classic channel error rate exactly by random sampling.
\item
Alice and Bob choose a proper LDPC code whose check matrix is $H$ according to the estimated error rate. Alice computes the syndrome $S_{1}$ of $X$, ${S_1} = H \times {X^T}$, where $X^{T}$ is the transpose of $X$ and sends $S_{1}$ to Bob.
\item
Bob receives $S_{1}$ and computes the syndrome $S_{2}$ of $Y$, ${S_2} = H \times {Y^T}$, where $Y^{T}$ is the transpose of $Y$. Bob compares $S_{1}$ and $S_{2}$. If $S_{1}$ is equal to $S_{2}$, the protocol ends successfully and Alice and Bob share a identical key string, otherwise, there are errors in $Y$ and Bob decodes his string with the help of $S_{1}$. If the syndrome of corrected $Y^{'}$ is equal to $S_{1}$, the protocol ends successfully. If the syndrome is still not equal to $S_{1}$, Bob sends a message to tell Alice and they can repeat the protocol if possible.

\end{enumerate}
Through error estimating step, Alice and Bob could choose a proper LDPC code to make this protocol success with high probability. We should note that, all two parities need to do to make protocol adaptive is done in advance, so they can easily implement this protocol without introducing extra computing in reconciliation period. Due to the inherent structure we use, we could store the base matrix with size $m\times n$ instead of mother matrix whose size is $mp\times np$ where $p$ is the size of sub-matrix, and this thus saves memory space significantly. Besides, because of the structure, decoder architecture is simplified and decoding is accelerated. These factors are important in practice.

\section{Simulation Results}

\begin{figure}[hbt]
\centering
\includegraphics[width=8cm,height=6cm]{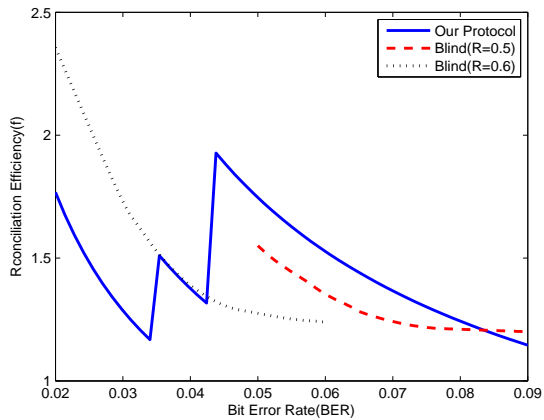}
\caption{Plot of efficiency versus Bit Error Rate. The efficiency of our scheme is compared to the efficiency of blind protocol. The efficiency of our scheme is calculated by using Eq.~\eqref{introduction:f2} and the efficiency of blind protocol is calculated according to \cite{martinez2012blind}.}
\label{Fig:efficiency1}
\end{figure}

\begin{figure}[hbt]
\centering
\includegraphics[width=8cm,height=6cm]{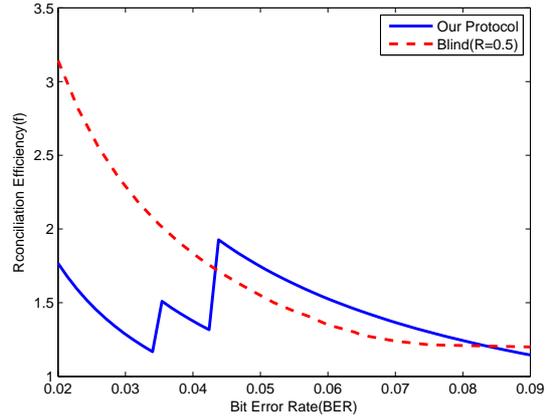}
\caption{Plot of efficiency versus Bit Error Rate. The performance of our scheme and blind protocol are compared when use just one LDPC code.}
\label{Fig:efficiency2}
\end{figure}

The simulation result is presented in Fig.~\ref{Fig:efficiency1}. In our simulation, we choose the code length as $1944$ bits which is short enough and easy to implement in practical. We compare the efficiency of our protocol and the efficiency of blind protocol proposed in \cite{martinez2012blind}. The degree distribution of mother code and effective codes are listed in Table~\ref{simulation:degreedistribution}. We set the size of sub-matrix as 54 which makes the maximum variable node degree large enough. The code rate of mother code is $1/2$, and the code rates of effective codes are $2/3$ and $3/4$. The code length of LDPC code used by blind protocol is $2000$ bits. Two LDPC codes are chosen for blind protocol to improve the performance and the code rate are $1/2$ and $3/5$. We set the maximum iteration of blind protocol to be $3$ and the proportion of punctured and shortened symbols is $10\%$. We should note that we just use one LDPC code(mother code) to cover the range of error rate. However, if we use one LDPC code in blind protocol, the performance will decrease, as shown in Fig.~\ref{Fig:efficiency2}.

From our simulation, we could find that our protocol works efficiently in the region of low error rate. Comparing to the blind protocol, our scheme performs better in low error rate region. However, there will be peaks in our scheme because when the error rate is too high to decode, code with lower code rate is chosen. To pull down the peaks, two possible ways are increasing the number of effective codes and increasing the performance of effective codes by setting stricter ACE criteria. Our scheme does't introduce extra computing while reconciling, so it is easily to implement.

\begin{table}[hbt]
%\centering
\caption{List of variable degree distribution of mother code and effective codes} \label{simulation:degreedistribution}
\begin{tabular}{c|lll}
\hline\hline
code rate &&& variable degree distribution\\
\hline
$1/2$ &&& $\lambda \left( x \right) = 0.2556{x^2} + 0.2932{x^3} + 0.4511{x^{10}}$\\
$2/3$ &&& $\lambda \left( x \right) = 0.176{x^2} + 0.456{x^3} + 0.168{x^7}$\\ &&& $+ 0.128{x^8} + 0.072{x^9}$\\
$3/4$ &&& $\lambda \left( x \right) = 0.129{x^2} + 0.5323{x^3} + 0.0968{x^6}$ \\ &&& $+ 0.1129{x^7} + 0.129{x^8}$\\
\hline\hline
\end{tabular}
\end{table}

\begin{figure}[hbt]
\centering
\includegraphics[width=8cm,height=6cm]{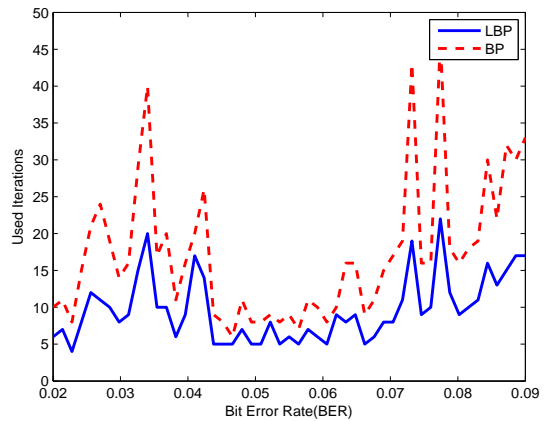}
\caption{Plot of used iterations of BP algorithm and LBP algorithm versus Bit Error Rate.}
\label{Fig:iteration}
\end{figure}

In simulation, we found the algorithm to design RCEV codes is hard to converge when there are too many effective codes. One possible way to solve this is to generate the highest code rate code first and the remaining effective codes can be generated by combining some of the rows chosen to generate the highest code rate code. We should note that, the generation of mother code and effective codes is preparing work and it won't slow down the reconciliation. We use both BP algorithm and LBP algorithm in simulation to decode codes. The iterations both algorithms used are compared in Fig.~\ref{Fig:iteration} and the result shows that LBP algorithm generally needs half of iterations BP algorithm used to decode. Therefore, LBP algorithm could accelerate decoding.

\section{Conclusion}

We propose an easily implemented rate compatible information reconciliation protocol based on RCEV LDPC codes. In our scheme, we use RCEV method to change the code rate. Simulation results show that our protocol works efficiently in the region of low error rate. Our scheme does't introduce extra computing to the reconciliation and thus can be easily implemented in practical. The codes we generate have inherent structure which could save memory space, simplify the decoder architecture and accelerate decoding process.

\section*{Acknowledgment}

This work is supported by National Natural Science Foundation of China Grant No.U1204602 and National High-Tech Program of China, Grant No.2011AA010803.

\end{document}